\documentclass[12pt]{article}
\usepackage{amssymb,setspace}
\usepackage{subeqn}
\usepackage{hyperref}

\newtheorem{theorem}{Theorem}
\newcommand{\ep}{\varepsilon}

\begin{document}

\title{Control of quantum transmission is trap free}
\author{Alexander N. Pechen$^{1,2}$\footnote{Corresponding author. E-mail:
\href{mailto:apechen@gmail.com}{apechen@gmail.com}; Webpage:
\href{http://mathnet.ru/eng/person17991}{mathnet.ru/eng/person17991}}\, and
David J. Tannor$^1$\\
\\
$^1$Department of Chemical Physics, Weizmann Institute
of Science\\
Rehovot 76100, Israel
\\
$^2$Steklov Mathematical Institute of Russian Academy of Sciences\\
Gubkina str. 8, Moscow 119991, Russia
}
\date{}
\maketitle

\begin{abstract}
We consider manipulation of the transmission coefficient for
a quantum particle moving in one dimension where the shape of the potential is
taken as the control. We show that the control landscape---the transmission as a
functional of the potential---has no traps, i.e., any maxima correspond to full
transmission.
\end{abstract}


{\it This article is part of a Special Issue dedicated to Professor Paul
Brumer in recognition of his contributions to chemistry.}

\section{Introduction}

Control of atomic and molecular scale systems obeying quantum equations of
motion is an important branch of modern science. Applications range from
selective excitation of atomic or molecular states to laser control of chemical
reactions and high harmonic
generation~\cite{Tannor1985,Rice2000,Brumer2003,Fradkov2003,Tannor2007,
Letokhov2007, Brif2012}.  One of the major
questions in quantum control theory is whether for a given objective the
control landscape
has traps, that is, local maxima with values less than the
global maximum~\cite{Rabitz2004,Pechen2011}. Much effort has been directed
towards the study of control landscapes of $n$-level systems. Despite this
effort, the proof of trap free behavior has been obtained so far only for
two-level systems~\cite{Pechen2012}. The case of
systems with an infinite dimensional Hilbert space has not been treated at all.

Here we consider control of transmission of a
quantum particle moving through a potential barrier where the shape of the
potential is used as a
control parameter. This is relevant, for example, to control of
tunneling~\cite{Brumer1999,Moiseev2002PRL,Moiseev2002IEEE}. We show
that the landscape of the transmission coefficient of a quantum particle as a
functional of the potential is trap free, i.e., any maxima correspond to full
transmission.

\section{Formulation}
Consider a particle of a fixed energy $E$ scattering on a barrier of potential
$V(x)$ which is assumed to have compact support ($V(x)=0$ when $|x|>a$ for some
$a$). The particle wavefunction satisfies the
time-independent Schr\"odinger equation
\begin{equation}\label{eq1}
 H_V\Psi(x)=E\Psi(x)
\end{equation}
where
\begin{eqnarray*}
 H_V=-\frac{d^2}{dx^2}+V(x) .
\end{eqnarray*}
We take the mass $m=1/2$ and $\hbar=1$.

The second-order differential equation~(\ref{eq1}) has two independent
solutions. We are free to choose linear combinations of the solutions that
behave as~\cite{Note}
\begin{subeqnarray}
\Psi^0_1(x)&=&
\left\{
\begin{array}{l}
e^{i k_E x}+A_Ee^{-i k_Ex},\quad x<-a\\
B_Ee^{i k_E x},\quad x>a
\end{array}\right.\label{eq2a}\\
\Psi^0_2(x)&=&
\left\{
\begin{array}{l}
D_Ee^{-i k_Ex},\quad x<-a\\
e^{-i k_E x}+C_Ee^{i
k_E x},\quad x>a
\end{array}\right. .\label{eq2b}
\end{subeqnarray}
Here $k_E=\sqrt{E}$. The solution $\Psi^0_1$ describes
the particle incident on
the barrier from
the left. The particle is partially reflected and partially transmitted
trough the barrier. Thus the wavefunction on
the left, far away from the barrier, is a sum of the incoming and reflected
waves, $\Psi^0_1(x)=e^{i k_Ex}+A_E e^{-i k_Ex}$ ($x\to -\infty$),
whereas on the
right of the barrier the wavefunction is an outgoing wave,
$\Psi^0_1(x)=B_Ee^{i k_Ex}$ ($x\to +\infty$). The coefficients $A_E$
and $B_E$ determine the probabilities of reflection and transmission,
respectively. The transmission coefficient is defined as the
amplitude of the transmitted wave, $T_E[V]=|B_E|^2$, and describes the
probability of transmission through the barrier. Similarly, the solution
$\Psi^0_2$
describes the particle incident on the barrier from
the right, which is partially reflected back to the right and partially
transmitted to the left~\cite{Ivanov2012}.

\subsection{Kinematic control landscape}
The general solution of Eq.~(\ref{eq1}) as $x\to -\infty$ is a sum of incoming
and
reflected waves $\Psi(x)=A'e^{i
k_Ex}+Ae^{-i k_Ex}$ and as $x\to +\infty$ is $\Psi(x)=B e^{i
k_Ex}+ B'e^{-i k_Ex}$. The coefficients $B$ and $B'$ are linearly related
to
the coefficients $A'$ and $A$ by a $2\times 2$ matrix $M$ which is
called the {\it monodromy operator}:
\begin{eqnarray*}
 \left(\begin{array}{c}
       B \\
       B'
      \end{array}\right)
=M
 \left(\begin{array}{c}
       A' \\
       A
      \end{array}\right)\, .
\end{eqnarray*}
The monodromy operator is an element of the {\it special
$(1,1)$ unitary group} $SU(1,1)$ also called the {\it real symplectic
group of second order} $Sp(1,\mathbb R)$~\cite{Arnold}. Any element of this
group can be represented as
\begin{eqnarray*}
M=
 \left(\begin{array}{cc}
       \sqrt{1+|z|^2}e^{i\phi} & z \\
       \bar z & \sqrt{1+|z|^2}e^{-i\phi}
      \end{array}\right)
\end{eqnarray*}
where $z\in\mathbb C$ and $\phi\in[0,2\pi)$.

Consider a wave incident on the potential from left infinity. Then $A'=1$,
$B'=0$
and the equality
\begin{eqnarray*}
 \left(\begin{array}{c}
       B \\
       0
      \end{array}\right)
=
 \left(\begin{array}{cc}
       M_{11} & M_{12} \\
       M_{21} & M_{22}
      \end{array}\right)
 \left(\begin{array}{c}
       1 \\
       A
      \end{array}\right)
\end{eqnarray*}
implies $A=-M_{21}/M_{22}$ and $B=1/M_{22}$. This gives for the transmission
coefficient (as a function of $M$) the {\it kinematic} expression
\begin{equation}\label{eq2}
 T(M)=|B|^2=\frac{1}{|M_{22}|^2}=\frac{1}{1+|z|^2}
\end{equation}

\begin{theorem}\label{theorem1}
The only extrema of $T(M)$ over $M\in SU(1,1)$ are global maxima. These occur
at $z=0$, where
\begin{eqnarray*}
M=
 \left(\begin{array}{cc}
       e^{i\phi} & 0 \\
       0 & e^{-i\phi}
      \end{array}\right),\qquad \phi\in[0,2\pi).
\end{eqnarray*}
\end{theorem}
{\bf Proof.} The theorem follows from eq.~(\ref{eq2}) and the domain of $z$.

Theorem~\ref{theorem1} shows that the control landscape of the transmission
coefficient has no kinematic traps and that its only kinematic extrema are
global maxima corresponding to full transmission.

\subsection{Dynamic control landscape}
What is of ultimate interest is to know if the {\it dynamic} landscape of the
transmission coefficient has traps, i.e. whether the transmission
coefficient as a functional
of the potential $V(x)$, has any local maxima or only a global maximum for full
transmission. In this section we
prove that there are no traps, i.e. all extrema of the transmission coefficient
$T_E[V]$ as a functional of the potential are only global maxima.

We will use the known result that for sufficiently smooth functions $f(E)$ and
$S(E)$
\begin{equation}
\int\limits_\mathbb R \frac{e^{ix
S(E_f)}f(E_f)}{E_f-E_i-i0}dE_f
= i\pi[1+{\rm sgn}\,S'(E_i)] f(E_i) e^{ix
S(E_i)}+O\left(x^{-\infty}\right)\label{lemma1}
\end{equation}
provided $S'(E_0)\ne 0$~\cite{Fedoruk1987,Avrimidi}.

\begin{theorem}
The only extrema of the objective $J[V]=T_E[V]$ are global maxima.
\end{theorem}
{\bf Proof.} Let $\Psi^0_{\alpha,E_i}(x)$ ($\alpha=1,2$) be two eigenfunctions
of $H_V$ with energy $E$. Consider a small variation of the potential $V(x)\to
V(x)+\delta v(x)$. The modification of the
eigenfunction with energy $E$ due to the variation of the potential can be
computed using perturbation theory for continuous
spectrum as follows (we omit a sum over the discrete spectrum since the
transmission coefficient depends only on the behavior of the wave function at
infinity, where wavefunctions corresponding to the discrete spectrum vanish)
\begin{equation}
\Psi_{1,E_i}=\Psi^0_{1,E_i}+\underbrace{\int\frac{\langle\Psi^0_{1,E_f}
, \delta
v\Psi^0_{1,E_i}\rangle}{E_i-E_f+i0}\Psi^0_{1,E_f}dE_f}\limits_{\delta\Psi_1(x)}
+\underbrace{\int\frac { \langle\Psi^0_ { 2 , E_f } , \delta
v\Psi^0_{1,E_i}\rangle}{E_i-E_f+i0}\Psi^0_{2,E_f}dE_f}\limits_{
\delta\Psi_2(x)}+o(\|\delta v\|)\label{eq5}
\end{equation}
Here $\langle\Psi^0_{\alpha,E_f},\delta v\Psi^0_{1,E_i}\rangle=\int_{\mathbb
R}\overline{\Psi}^0_{\alpha,E_f}(x)\delta v(x)\Psi^0_{1,E_i}(x)dx$ for
$\alpha=1,2$.

The transmission coefficient at energy $E_i$ for the modified potential
$V+\delta v$ up to linear order in $\delta v$ can be computed as
\begin{eqnarray*}
T_{E_i}[V+\delta v]&=&\lim\limits_{x\to +\infty}|\Psi_{1,E_i}(x)|^2\\
&=&\lim\limits_{x\to+\infty}\Bigl\{|\Psi^0_{1,E_i}
(x)|^2 +2\Re\Bigl[\overline{\Psi}^0_{1,E_i}
(x)\Bigl(\delta\Psi_1(x)+\delta\Psi_2(x)\Bigr)\Bigr]\Bigr\}
+o(\|\delta v\|)\\
&=&T_{E_i}[V]+\delta J(V)+o(\|\delta v\|)
\end{eqnarray*}
By eqs.~(\ref{eq2a}),~(\ref{lemma1}), and~(\ref{eq5})
\begin{eqnarray*}
\lim\limits_{x\to+\infty}2\Re[\overline{\Psi}^0_{1,E_i}
(x)\delta\Psi_1(x) ]&=&-2\Re\lim\limits_{x\to
+\infty}\int\frac{\langle\Psi^0_{1,E_f},\delta
v\Psi^0_{1,E_i}\rangle}{E_f-E_i-i0}B^*_{E_i} B_{E_f}e^{i
(k_{E_f}-k_{E_i})x}\\
&=&-4\pi\Im[\langle\Psi^0_{1,E_i},\delta v\Psi^0_{1,E_i}\rangle|B_{E_i}|^2]=0
\end{eqnarray*}
Here the second equality follows from~(\ref{lemma1}) with
$S(E_f)=k_{E_f}-k_{E_i}$ which gives ${\rm sgn}\, S'(E_f)|_{E_f=E_i}=1$, and the
last equality follows from the fact that diagonal matrix elements of $\delta v$
are real. Similarly,
\begin{eqnarray*}
\lefteqn{\lim\limits_{x\to+\infty}2\Re[\overline{\Psi}^0_{1,E_i}
(x)\delta\Psi_2(x) ]} \\
&=&-2\Re\lim\limits_{x\to+\infty} \int\frac{\langle\Psi^0_{2,E_f},\delta
v\Psi^0_{1,E_i}\rangle}{E_f-E_i-i0}B^*_{E_i}e^{-i
k_{E_i}x}[e^{-i k_{E_f}x}+C_{E_f}e^{i k_{E_f}x}]dE_f\\
&=&-2\Re\lim\limits_{x\to+\infty} \int\frac{\langle\Psi^0_{2,E_f},\delta
v\Psi^0_{1,E_i}\rangle}{E_f-E_i-i0}B^*_{E_i}[e^{-i
(k_{E_f}+k_{E_i})x}+C_{E_f}e^{i
(k_{E_f}-k_{E_i})x}]dE_f
\end{eqnarray*}
The term with $e^{-i (k_{E_f}+k_{E_i})x}$ in the square brackets gives
zero contribution in the limit since for this
term $S(E_f)=-k_{E_f}-k_{E_i}$ gives ${\rm sgn}\, S'(E_f)=-1$ and the integral
is
$O(x^{-\infty})$ according to~(\ref{lemma1}). The contribution of the term with
$e^{i (k_{E_f}-k_{E_i})x}$ in the square brackets can be computed using
equality~(\ref{lemma1}) as follows:
\begin{eqnarray*}
-2\lim\limits_{x\to+\infty}\Re\int\frac{\langle\Psi^0_ {2,E_f} ,
\delta
v\Psi^0_{1,E_i}\rangle}{E_f-E_i-i0}B^*_{E_i}C_{E_f}e^{i
(k_{E_f}-k_{E_i})x}dE_f
&=&-4\pi\Re\left[i\langle\Psi^0_{2,E_i},\delta
v\Psi^0_{1,E_i}\rangle
B^*_{E_i}C_{E_i}\right]\vphantom{\frac{A^*_{E_i}}{B^*_{E_i}}}\\
&=&-4\pi |B_{E_i}|^2\Im\left[\langle\Psi^0_{2,E_i},\delta
v\Psi^0_{1,E_i}\rangle\frac{A^*_{E_i}}{B^*_{E_i}}\right]\\
&=&-4\pi T_{E_i}[V]\Im\left[\langle\Psi^0_{2,E_i},\delta
v\Psi^0_{1,E_i}\rangle\frac{A^*_{E_i}}{B^*_{E_i}}\right]\\&=&\delta J(V)
\end{eqnarray*}
Here we have used the fact for any $E$
\[
 C(E)=-\frac{B(E) A^*(E)}{B^*(E)}
\]
(see Eqs.~(7.84)
and~(7.86) in~\cite{Tannor2007}).

A critical potential $V(x)$ should satisfy $\delta J(V)=0$ for any
$\delta v$. Since for any $E$ $B(E)\ne 0$ and $T_E[V]\ne 0$, this is possible
only if
$A(E_i)=0$. That corresponds to $T_{E_i}[V]=1$, i.e., any critical potential
leads to a global maximum of the transmission coefficient. This concludes the
proof of the theorem.

\section{Comparison with the landscape for coherent control by lasers}

Quantum control landscapes for $n$-level systems controlled by lasers or
electro-magnetic fields have been extensively studied in recent years. In this
section,
we put our findings about the landscape for
control of transmission in the context of what is known about the landscape for
coherent
control of $n$-level systems by lasers. We assume that the $n$-level system
interacts only with the laser and is isolated from other environments, i.e., is
a
closed quantum system.

The evolution equation for a system controlled by a laser field $\ep(t)$ is the
Schr\"odinger equation
\[
 i\frac{dU^\ep_t}{dt}=(H_0+\ep(t)V)U^\ep_t,\qquad U^\ep_0=\mathbb I
\]
Here $H_0$ and $V$ are the free and interaction Hamiltonians,
respectively. The solution of this equation is a unitary matrix, $U^\ep_T\in
U(n)$.
The overall phase of the unitary evolution operator is physically meaningless,
so
that $U$ and $e^{i\phi}U$ describe the same physics. Thus the space of kinematic
controls for laser control is the special unitary group $SU(n)$, instead of the
special $(1,1)$ unitary group $SU(1,1)$ for control of the transmission
coefficient. The objective that corresponds to the transmission coefficient is
the probability of transition from some initial state $\psi_i$ to some final
state $\psi_f$, $J(U_T^\ep)=|\langle\psi_f|U^\ep_T|\psi_i\rangle|^2$.

The kinematic landscape
for the transition probability $J(U)$ (considered as a function of $U\in SU(n)$)
is trap-free~\cite{Rabitz2004}. However, this result does not imply the absence
of traps for the corresponding dynamic landscape $J(U^\ep_T)$ (considered as a
functional of $\ep$) due to the existence of non-regular controls (i.e. controls
where the rank of the
Jacobian $\delta U_T/\delta\ep(t)$ is not maximal). To see this, consider the
chain rule
\[
\frac{\delta J(\ep)}{\delta\ep(t)}=\frac{\delta J(U)}{\delta
U}\biggl|_{U=U^\ep_T} \frac{\delta U^\ep_T}{\delta\ep(t)}.
\]
The absence of traps for $J(U)$ implies the absence of traps for $J(\ep)$ only
if the Jacobian $\delta U^\ep_T/\delta\ep(t)$ has full rank, i.e. has no zero
eigenvalues. The analogous full rank criterion for control of the transmission
coefficient
is that rank of the Jacobian $\delta M_V/\delta V(x)$ is maximal, where $M_V$ is
the monodromy operator for potential $V(x)$.

The only
rigorous proof of the absence of dynamical traps for coherent laser control is
for
$n=2$~\cite{Pechen2012}. Interestingly, the dimension of the corresponding
kinematic control space $SU(2)$ is the same as
the dimension of the kinematic control space $SU(1,1)$ for control of
transmission. While the resulting conclusion of trap-free dynamics is the same
for these two cases, the proofs are fundamentally different.

We summarize the comparison of landscape-related notions for laser control
and for control of transmission in Table~1.

\begin{table}\label{table1}
\begin{tabular}{|p{4.2cm}|p{5.5cm}|p{5.5cm}|}
\hline
& Coherent control by laser & Control by potential  \tabularnewline
\hline
Dynamic control & Laser pulse $\ep(t)$ & Potential $V(x)$  \tabularnewline
\hline
Kinematic control & $U\in SU(n)$ & $M\in SU(1,1)$  \tabularnewline
\hline
Objective & $J(U)=|\langle\psi_f|U|\psi_i\rangle|^2$ &
$J(M)=\vphantom{\int\limits_A^B} \frac{\textstyle 1}{\textstyle |M_{22}|^2}$
\tabularnewline
\hline
Kinematic landscape & No traps, & No traps,  \tabularnewline
& ${\rm max}\, J=1$, ${\rm min}\, J=0$ & ${\rm max}\, J=1$, ${\rm inf}\, J=0$
\tabularnewline
\hline
Full rank criterion & Jacobian $\delta U^\ep_T/\delta\ep(t)$ has maximal rank
&
Jacobian $\delta M_V/\delta V(x)$ has maximal rank \tabularnewline
\hline
Dynamic landscape & Generally unknown. Trap-free for $n=2$. & Trap-free.
\tabularnewline
\hline
\end{tabular}
\caption{Comparison of landscape-related notions for coherent control by
lasers and
control by potential.}
\end{table}

\section*{Acknowledgments}
A.N. Pechen acknowledges support of the Marie Curie International Incoming
Fellowship within the 7th European Community Framework Programme. This
research is made possible in part by the historic generosity of the Harold
Perlman family.

\end{document}